\documentclass[10pt,aps,twocolumn,secnumarabic,balancelastpage,amsmath,amssymb,nofootinbib,hyperref=pdftex,prx,citeautoscript,showpacs]{revtex4-1}

\usepackage{graphics}
\usepackage[pdftex]{graphicx}
\usepackage{tabularx}
\graphicspath{{./Figures/}}
\usepackage{longtable}
\usepackage{amsmath}
\usepackage{color,soul}
\usepackage{braket}

\makeatletter
\def\p@subsection{\thesection\,}
\makeatother

\usepackage{hyperref}
\hypersetup{colorlinks=true,linkcolor=blue,anchorcolor=blue,citecolor=blue,filecolor=blue,urlcolor=blue,bookmarksnumbered=true,pdfview=FitB}

\begin{document}

\title{Finite-size effects in a nanowire strongly coupled to a thin superconducting shell}
\author{Christopher Reeg}
\author{Daniel Loss}
\author{Jelena Klinovaja}
\affiliation{Department of Physics, University of Basel, Klingelbergstrasse 82, CH-4056 Basel, Switzerland}
\date{\today}
\begin{abstract}
We study the proximity effect in a one-dimensional nanowire strongly coupled to a finite superconductor with a characteristic size which is much shorter than its coherence length. Such geometries have become increasingly relevant in recent years in the experimental search for Majorana fermions with the development of thin epitaxial Al shells which form a very strong contact with either InAs or InSb nanowires. So far, however, no theoretical treatment of the proximity effect in these systems has accounted for the finite size of the superconducting film. We show that the finite-size effects become very detrimental when the level spacing of the superconductor greatly exceeds its energy gap. Without any fine-tuning of the size of the superconductor (on the scale of the Fermi wavelength), the tunneling energy scale must be larger than the level spacing in order to reach the ``hard gap" regime which is seen ubiquitously in the experiments. However, in this regime, the large tunneling energy scale induces a large shift in the effective chemical potential of the nanowire and pushes the topological phase transition to magnetic field strengths which exceed the critical field of Al.
\end{abstract}
\pacs{74.45.+c,71.10.Pm,73.21.Hb,74.78.Na}

\maketitle

\section{Introduction} Topological superconductivity has been a subject of intense study in recent years \cite{Alicea:2012} both theoretically  and experimentally because the localized Majorana excitations of such systems obey non-Abelian statistics and can potentially be utilized for applications in quantum computing \cite{Kitaev:2001,Nayak:2008}. The most promising proposal to date for engineering Majorana bound states in nanowires combines Rashba spin-orbit coupling, proximity-induced $s$-wave superconductivity, and an external magnetic field applied parallel to the nanowire \cite{Lutchyn:2010,Oreg:2010,Mourik:2012,Deng:2012,Das:2012,Rokhinson:2012,Churchill:2013,Finck:2013,Chang:2015,Albrecht:2016,Deng:2016,Zhang:2017,Gazibegovic:2017}. An alternative proposal which has also received a great deal of attention involves coupling a ferromagnetic atomic chain to an $s$-wave superconductor with strong intrinsic spin-orbit coupling  \cite{NadjPerge:2014,Ruby:2015,Pawlak:2016,Klinovaja:2013,Vazifeh:2013,Braunecker:2013,NadjPerge:2013,Pientka:2013,Awoga:2017}. Since the first generation of nanowire experiments \cite{Mourik:2012,Deng:2012,Das:2012,Rokhinson:2012,Churchill:2013,Finck:2013}, there has been significant progress made in both the fabrication of cleaner devices as well as in the quality of the proximity-induced superconductivity \cite{Chang:2015,Albrecht:2016,Deng:2016,Zhang:2017,Gazibegovic:2017}. The most significant advance in this respect has been the development of thin shells (with thickness $d\sim10$ nm) of superconducting Al grown epitaxially on either InAs \cite{Chang:2015,Albrecht:2016,Deng:2016} or InSb \cite{Gazibegovic:2017} nanowires, thus ensuring a very strong proximity contact which has led to very hard induced gaps in the nanowires in the absence of a magnetic field.

Despite the recent experimental development of these thin superconducting shells, the most comprehensive theories describing proximity-induced superconductivity in a nanowire treat the superconductor as infinitely large \cite{Sau:2010prox,Potter:2011,Kopnin:2011,Zyuzin:2013,Deng:2016,vanHeck:2016,Reeg:2017_2}. Such an assumption implies that there is a continuum of states in the superconductor, and therefore there are always states available to couple to the nanowire and open a gap. However, in reality, the superconductor has a finite level spacing $\delta E_s\sim \hbar v_F/d$ due to its finite size. For the thin Al shells studied experimentally ($v_F\sim10^6$ m$/$s and $d\sim10$ nm), the level spacing of the shell $\delta E_s\sim10$ meV exceeds the Fermi energy of the nanowire ($\sim0.1-1$ meV for typical semiconducting nanowires). Thus, for the experimental system, the limit of a bulk superconductor is not the relevant one and finite-size effects are expected to play an important role in determining the strength of proximity-induced superconductivity.

In this paper, we show that the finite size of the shell can be very detrimental to inducing superconductivity in the nanowire. In order to induce a sizable superconducting gap without finely tuning the thickness of the shell on the scale of the Fermi wavelength (of the superconductor), the energy scale describing tunneling between the nanowire and superconductor ($\gamma$) must be made larger than the level spacing of the shell ($\gamma\gtrsim \delta E_s$). However, such strong tunneling induces a shift in the effective chemical potential of the nanowire which greatly exceeds the semiconducting energy scale. As a result, it is possible for the system to exhibit a hard gap even if the nanowire is effectively depleted; in this case, the gap is determined by the lowest subband in the superconductor rather than the nanowire. Additionally, in order to reach the topological phase, the Zeeman energy induced by an external magnetic field must counteract the large chemical potential shift. As a result, the field strength needed to reach the topological phase greatly exceeds the critical field of Al.

The remainder of this paper is organized as follows. In Sec.~\ref{Sec2}, we describe a simple theoretical model which can be applied to the experimental geometry of a thin superconducting shell strongly coupled to a semiconducting nanowire. In Sec.~\ref{Sec3}, we analyze the spectrum of our model, showing that a large tunneling strength is needed to overcome the large level spacing of the superconductor and open a sizable gap in the nanowire. We consider the case when the nanowire is located near the edge of the superconductor in Sec.~\ref{Sec3a}, while we consider the case when the nanowire is located in the middle of the superconductor in Sec.~\ref{Sec3b}. We present a numerical tight-binding calculation to back up the theoretical analysis of our model in Sec.~\ref{Sec4}. In Sec.~\ref{Sec5}, we determine how the finite size of the superconductor affects the critical field strength needed to reach the topological phase in the nanowire. In Sec.~\ref{Sec6}, we relate the results of our simple model directly to the experimental setup and provide estimates for the level spacing of the superconducting shell, the tunneling strength needed to induce a sizable gap in the nanowire, and the critical field strength needed to reach the topological phase. Our conclusions are given in Sec.~\ref{Sec7}.

\section{Model} \label{Sec2}
\begin{figure}[t!]
\includegraphics[width=\linewidth]{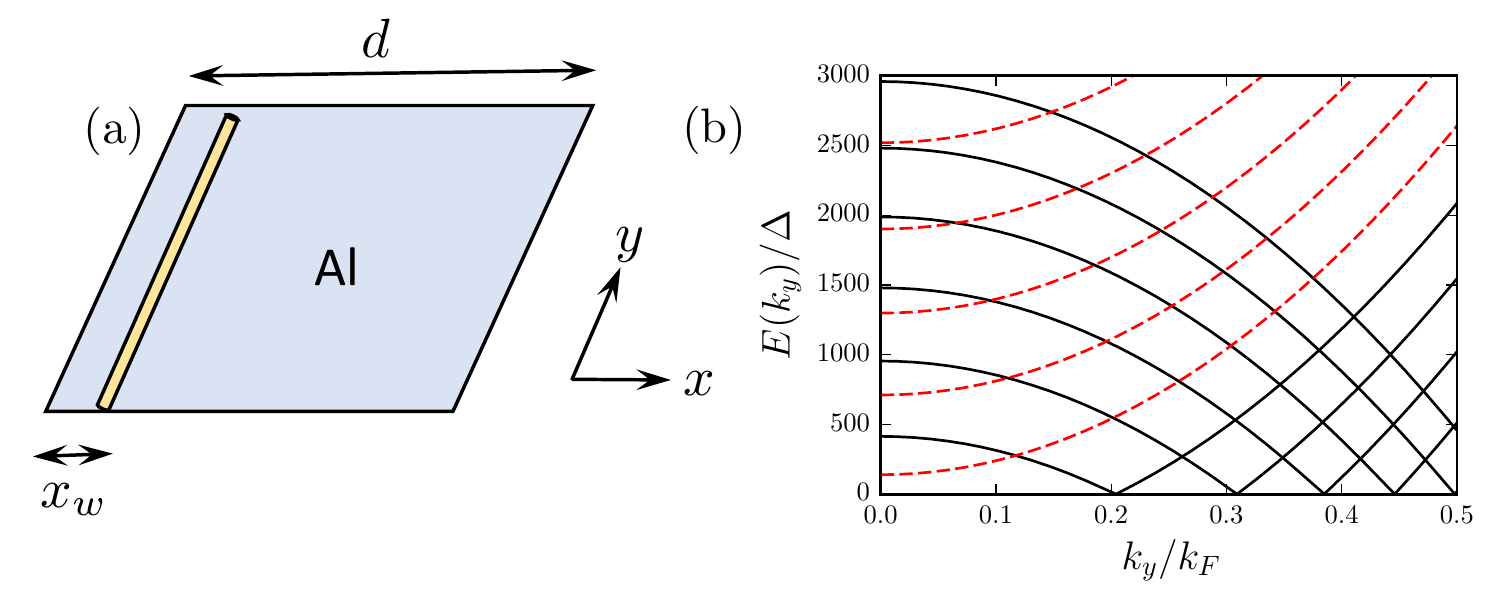}
\caption{\label{setup}(a) A single-channel 1D nanowire is tunnel-coupled (at position $x_w$) to a superconductor with finite extent $d$ in the direction perpendicular to the nanowire. (b) Spectrum of finite-sized superconductor in the absence of tunneling [Eq.~\eqref{barespec}] with $\mu_s/\Delta=10^4$ and $k_Fd/\pi=35.75$. Each occupied subband (black) has a gap of $\Delta$ which is not visible on the scale of the plot; unoccupied subbands (red) do not have a superconducting gap. The subband spacing in this case is much larger than both the superconducting gap and the characteristic energy scale of the nanowire, $\delta E_s\gg\Delta,\mu_w$.}
\end{figure}

The system we consider is displayed in Fig.~\ref{setup}(a). We consider a nanowire which is an infinitely long one-dimensional channel oriented along the $y$-direction (with zero width). The nanowire is tunnel coupled at a position $x=x_w$ to a superconductor which is infinitely long in the $y$-direction and has finite extent $d$ in the $x$-direction (the need for a finite $x_w$ will be explained below).

We consider a Hamiltonian of the form
\begin{equation} \label{Hamiltonian}
H=H_{w}+H_{s}+H_t.
\end{equation}
For now, we take a simple model for the Hamiltonian of the nanowire,
\begin{equation} \label{Hnw}
H_{w}=\sum_{\sigma}\int\frac{dk_y}{2\pi}\,\psi^\dagger_{\sigma}(k_y)\xi_k\psi_{\sigma}(k_y),
\end{equation}
where $k_y$ is a conserved momentum in the direction parallel to the nanowire, $\psi_{\sigma}^\dagger(k_y)$ is the creation operator in the nanowire, and $\xi_k=k_y^2/2m_w-\mu_w$ ($m_w$ and $\mu_w$ are the effective mass and chemical potential of the nanowire, respectively). The superconductor is described by a BCS Hamiltonian,
\begin{equation} \label{Hs}
\begin{aligned}
H_{s}=\frac{1}{2}\int\frac{dk_y}{2\pi}\int_0^d dx\,\eta^\dagger(k_y,x)\mathcal{H}_{\text{BCS}}\eta(k_y,x),
\end{aligned}
\end{equation}
where $\eta(k_y,x)=[\eta_{\uparrow}(k_y,x),\eta_{\downarrow}^\dagger(-k_y,x)]^T$, $\eta_{\sigma}^\dagger(k_y,x)$ is the creation operator in the superconductor, and $\mathcal{H}_\text{BCS}=(-\partial_x^2/2m_s+k_y^2/2m_s-\mu_s)\tau_z+\Delta\tau_x$, with $\Delta$ the (constant in space) superconducting pairing potential and $\tau_{x,y,z}$ Pauli matrices acting in Nambu space. The two systems are coupled at a position $x=x_w$ by a tunneling term which we assume preserves spin and momentum,
\begin{equation} \label{Ht}
H_t=-t\sum_\sigma \int\frac{dk_y}{2\pi}\left[\psi_{\sigma}^\dagger(k_y)\eta_{\sigma}(k_y,x_w)+H.c.\right],
\end{equation}
where $t$ is the (spin-independent) tunneling amplitude. Such a model corresponds to local tunneling along the superconductor/semiconductor interface.

In the absence of tunneling, the spectrum of the finite-sized superconductor is given by ($n\in\mathbb{Z}^+$)
\begin{equation} \label{barespec}
E_n(k_y)=\sqrt{\left(\mu_s-\frac{k_y^2}{2m_s}-\frac{\pi^2 n^2}{2m_sd^2}\right)^2+\Delta^2}.
\end{equation}
When the quantization scale exceeds the gap, $1/m_sd^2\gg\Delta$, there are very few subbands available to couple to the low-energy modes of the nanowire, as shown in Fig.~\ref{setup}(b). In this case, the relevant subbands follow a linearized form
\begin{equation} \label{barespec2}
E_n(k_y)=\sqrt{\left[(k_Fd-\pi n)\delta E_s-\frac{k_y^2}{2m_s}\right]^2+\Delta^2},
\end{equation}
where we define the level spacing $\delta E_s=v_F/d$ ($v_F=k_F/m_s$ is the Fermi velocity of the superconductor and $k_F=\sqrt{2m_s\mu_s}$ is the Fermi momentum). As we will show explicitly, when $\delta E_s\gg\Delta$, it is the level spacing which is the relevant scale (rather than $\Delta$) in determining the strength of the proximity effect.

The Hamiltonian in Eq.~\eqref{Hamiltonian} can be diagonalized by means of a Bogoliubov transformation \cite{Reeg:2017}. The resulting Bogoliubov-de Gennes (BdG) equation is given by
\begin{equation} \label{BdG}
\left[\mathcal{H}_\text{BCS}+t^2\delta(x-x_w)G_0^R(E,k_y)\right]\psi_s(x)=E\psi_s(x),
\end{equation}
where $\psi_s(x)$ is the (Nambu spinor) wave function of the superconductor and $G_0^R(E,k_y)=(E-\xi_k\tau_z+i0^+)^{-1}$ is the bare retarded Green's function of the nanowire (in the absence of tunneling). The nanowire itself enters only through the boundary condition at $x=x_w$ [corresponding to the $\delta$-function term in Eq.~\eqref{BdG}]. Inside the superconductor, we solve Eq.~\eqref{BdG} on both the left ($x<x_w$) and right ($x>x_w$) sides of the nanowire to give
\begin{subequations} \label{WF}
\begin{gather}
\psi_L(x)=c_1\left(\begin{array}{c} u_0 \\ v_0 \end{array}\right)\sin(k_+d)+c_2\left(\begin{array}{c} v_0 \\ u_0 \end{array}\right)\sin(k_-d), \\
\psi_R(x)=c_3\left(\begin{array}{c} u_0 \\ v_0 \end{array}\right)\sin[k_+(d-x)]+c_4\left(\begin{array}{c} v_0 \\ u_0 \end{array}\right)\sin[k_-(d-x)],
\end{gather}
\end{subequations}
where $k_\pm=(k_F^2-k_y^2\pm2im_s\Omega)^{1/2}$, $u_0(v_0)=\sqrt{(1\pm i\Omega/E)/2}$, and $\Omega=\sqrt{\Delta^2-E^2}$. The vanishing boundary conditions which we impose at the free ends of the superconductor ($x=0$ and $x=d$) are accounted for already in Eqs.~\eqref{WF}; the boundary conditions at $x=x_w$ due to tunneling are given by \cite{Reeg:2017}
\begin{subequations} \label{BCs}
\begin{gather}
\psi_L(x_w)=\psi_R(x_w), \\
\frac{1}{k_F}\bigl[\partial_x\psi_R(x_w)-\partial_x\psi_L(x_w)\bigr]=2\gamma\tau_zG_0^R(E,k_y)\psi_s(x_w),
\end{gather}
\end{subequations}
where $\gamma=t^2/v_F$ is a tunneling energy scale.

Imposing boundary conditions at $x=x_w$, the solvability condition of the resulting system of equations determines the excitation spectrum $E(k_y)$. Assuming that $\mu_s\gg|\Omega|$, we make a semiclassical expansion
\begin{equation}
k_\pm=k_F\varphi\pm i\Omega/(v_F\varphi)\equiv \zeta\pm i\chi,
\end{equation}
where $\varphi=(1-k_y^2/k_F^2)^{1/2}$ parametrizes the quasiparticle trajectory inside the superconductor ($0<\varphi\leq1$). We note that the semiclassical approximation breaks down for grazing trajectories $k_y\approx k_F$ within the superconductor, and we do not consider such trajectories. After some algebra (see Appendix~\ref{appA} for details), the solvability condition can be expressed as
\begin{equation} \label{solvability}
\frac{E^2}{\Gamma^2(E,k_y)}-\Delta^2\left(\frac{1}{\Gamma(E,k_y)}-1\right)^2-[\xi_k-\delta\mu(E,k_y)]^2=0,
\end{equation}
where we define the effective parameters
\begin{widetext}
\begin{gather}
\Gamma=\biggl(1+\frac{\gamma}{\Omega\varphi[\cosh(2\chi d)-\cos(2\zeta d)]}\biggl\{\sinh(2\chi d)-\cos(2\zeta x_w)\sinh[2\chi(d-x_w)]-\cos[2\zeta(d-x_w)]\sinh(2\chi x_w)\biggr\}\biggr)^{-1}, \nonumber \\
\delta\mu=-\frac{\gamma}{\varphi[\cosh(2\chi d)-\cos(2\zeta d)]}\biggl\{\sin(2\zeta d)-\sin(2\zeta x_w)\cosh[2\chi(d-x_w)]-\sin[2\zeta(d-x_w)]\cosh(2\chi x_w)\biggr\}. \label{effparams}
\end{gather}
\end{widetext}
The quantity $\Gamma(E,k_y)$, which takes values $0<\Gamma<1$ for $E<\Delta$, renormalizes the energy and is responsible for inducing superconductivity in the nanowire, while the quantity $\delta\mu(E,k_y)$ corresponds to a tunneling-induced shift in the effective chemical potential of the nanowire.

Note that both tunneling-induced terms vanish ($\delta\mu=0$ and $\Gamma=1$) if the nanowire is taken to be strictly at the edge of the superconductor, $x_w=0$ or $x_w=d$. As a result, the system behaves as though there is no tunnel coupling [i.e., Eq.~\eqref{solvability} reduces to simply $E^2=\xi_k^2$]. This is a direct consequence of the fact that the nanowire was taken to have zero width. The tunneling term in Eq.~\eqref{BdG} relates the wave functions in the nanowire and superconductor at $x=x_w$, and the superconducting wave function vanishes at the boundaries. We choose to keep the approximation of a zero-width nanowire, as it is more consistent with previous related theories and is easier to treat analytically, and therefore the nanowire must be chosen to be located at some position $x=x_w\neq0$ in order to have a non-vanishing tunnel coupling. We will show in Sec.~\ref{Sec4} that such an approximation is consistent with a numerical tight-binding calculation in which the wire can be placed strictly at the edge of the superconductor. Alternatively, if the nanowire is located strictly at the edge of the superconductor, it must be given a finite width so that the superconducting wave function does not vanish at the interface. Related calculations were carried out in Refs.~\cite{Volkov:1995,Fagas:2005,Tkachov:2005,Reeg:2016}, where proximity-induced superconductivity was studied in a quasi-two-dimensional layer of finite width coupled to a semi-infinite three-dimensional superconductor.

We also note that we can equivalently express the solvability condition in Eq.~\eqref{solvability} in the language of Green's functions. We can rewrite Eq.~\eqref{solvability} in the form 
$\det(G^R_w)^{-1}=0$, where 
 $ G^R_w=[({G}_0^R)^{-1}-\Sigma^R]^{-1}$
is the retarded Green's function of the nanowire with a self-energy induced by the superconductor. From Eq.~\eqref{solvability}, we can identify the retarded self-energy as (see also Appendix~\ref{appA})
\begin{equation} \label{selfenergy}
\Sigma^R(E,k_y)=(1/\Gamma-1)(\Delta\tau_x-E)-\delta\mu\,\tau_z.
\end{equation}
with $\Gamma$ and $\delta\mu$ as defined in Eq.~\eqref{effparams}.

Before moving on, we pause to compare our result for the self-energy of a nanowire coupled to a finite-sized superconductor to the self-energy that has appeared extensively in the literature to describe proximitized nanowires beyond the weak coupling limit \cite{Deng:2016,Sau:2010prox,Potter:2011,Kopnin:2011,Zyuzin:2013,vanHeck:2016,Reeg:2017_2}. In these works, all based on the approach of integrating out the superconducting degrees of freedom, the superconductor is implicitly assumed to be infinitely large, with a nanowire coupled to the middle of the superconductor. In this geometry, one obtains the same self-energy as given in Eq.~\eqref{selfenergy}, but with the vastly simplified effective parameters $\delta\mu=0$ and $\Gamma=(1+\gamma/\Omega)^{-1}$. We find that we recover this form for the self-energy by setting $x_w=d/2$ and taking the limit $d\to\infty$ in Eq.~\eqref{effparams} (the momentum dependence must also be neglected by setting $\varphi=1$). For maximum transparency in relating the current work to the previous ones, we show in Appendix~\ref{appB} how Eqs.~\eqref{effparams} and \eqref{selfenergy} can be equivalently derived by integrating out the superconductor.

\section{Excitation Spectrum} \label{Sec3}
In this section, we analyze the excitation spectrum of our model in two simplified limiting cases. First, in Sec.~\ref{Sec3a}, we consider the case when the nanowire is placed very close to the boundary of the superconductor, such that $k_Fx_w\ll1$ (i.e., the distance between the nanowire and the edge of the system is much smaller than the Fermi wavelength of the superconductor $\lambda_F$). In Sec.~\ref{Sec3b}, we consider the case when the nanowire is placed in the middle of the superconductor, $x_w=d/2$. Throughout this section, we assume that the width of the superconductor is much smaller than its coherence length, $d\ll\xi_s$; equivalently, its level spacing is much larger than the gap, $\delta E_s\gg\Delta$.

\subsection{Wire near edge of superconductor} \label{Sec3a}
\subsubsection{Analytical calculation of excitation gap} \label{Sec3a1}
We first look to analytically determine the excitation gap of the semiconductor/superconductor system when the wire is placed near the edge of the superconductor ($k_Fx_w\ll1$). In this limit, and taking $E<\Delta$ ($\Delta$ is the upper bound on the size of the excitation gap), the effective parameters of Eq.~\eqref{effparams} can be simplified to
\begin{equation} \label{effparamsend}
\begin{gathered}
\Gamma=\left(1+\frac{2\gamma(k_Fx_w)^2}{\delta E_s\sin^2(k_Fd)}\right)^{-1}, \\
\delta\mu=2\gamma(k_Fx_w)[1-(k_Fx_w)\cot(k_Fd)].
\end{gathered}
\end{equation}
In Eq.~\eqref{effparamsend}, we have additionally assumed that $|\sin(k_Fd)|\gg\Delta/\delta E_s$ (recall that we are assuming $\Delta/\delta E_s\ll1$). Therefore, Eq.~\eqref{effparamsend} breaks down when the thickness of the shell approaches $k_Fd\to \pi n$ ($n\in\mathbb{Z}^+$). We have also neglected the momentum dependence of $\varphi(k_y)$ by setting $\varphi=1$; this assumption is justified provided that $k_y/k_F\ll1/\sqrt{k_Fd}$.

Because the effective parameters of Eq.~\eqref{effparamsend} are not functions of energy or momentum, it is particularly simple to solve for the spectrum,
\begin{equation} \label{specend}
E^2=\Gamma^2\left(\frac{k_y^2}{2m_w}-\mu_\text{eff}\right)^2+\Delta^2(1-\Gamma)^2,
\end{equation}
where we define $\mu_\text{eff}=\mu_w+\delta\mu$. [Remember, Eq.~\eqref{specend} should be taken to describe the spectrum only for $E<\Delta$.] If $\mu_\text{eff}>0$, the spectrum of Eq.~\eqref{specend} describes a superconductor with an induced gap of size
\begin{equation} \label{gapend}
E_g/\Delta=1-\left(1+\frac{2\gamma(k_Fx_w)^2}{\delta E_s\sin^2(k_Fd)}\right)^{-1}
\end{equation}
which is opened around the effective Fermi momentum $k_{F,\text{eff}}=\sqrt{2m_w\mu_\text{eff}}$. We note that Eq.~\eqref{gapend} cannot be applied if the tunneling energy is made too large, such that $k_{F,\text{eff}}/k_F\gtrsim1/\sqrt{k_Fd}$. In terms of energy scales, we find that Eq.~\eqref{gapend} breaks down when $\sqrt{\gamma/\delta E_s}\gtrsim\sqrt{(m_s/m_w)/(k_Fx_w)}\gg1$. If $\mu_\text{eff}<0$, then Eq.~\eqref{specend} describes the spectrum of an insulator with gap $|\mu_\text{eff}|$. In this case, one must take into account the full momentum dependence $\varphi(k_y)$ in order to calculate the gap and Eq.~\eqref{gapend} does not apply.

In the limit when the induced gap is small $E_g\ll\Delta$, it is necessarily given by
\begin{equation} \label{gapsmall}
E_g/\Delta=\frac{2\gamma(k_Fx_w)^2}{\delta E_s\sin^2(k_Fd)}\ll1.
\end{equation}
This result has several important implications. First, assuming that the thickness of the shell is not finely tuned on the scale of the Fermi wavelength of the superconductor [i.e., $\sin^2(k_Fd)\sim1$], we see that the induced gap can be small even if the tunneling energy greatly exceeds the gap of the superconductor ($\gamma\gg\Delta$). This result is purely a finite-size effect and is due to the suppression of the gap by a factor $\Delta/\delta E_s\ll1$. Second, we see that the gap is additionally suppressed by a factor $(k_Fx_w)^2\ll1$, which is a direct consequence of the smallness of the superconducting wave function in the vicinity of the edge. Finally, we note that it is still possible to induce a sizable gap if the thickness of the shell is fine-tuned to the limit $\sin^2(k_Fd)\ll(\gamma/\delta E_s)(k_Fx_w)^2\ll1$; in this limit, we find from Eq.~\eqref{gapsmall} that $E_g\gg\Delta$ and our original expansion breaks down. This leads to a resonance behavior, with sharp peaks in the induced gap when the resonance condition $\sin(k_Fd)=0$ is satisfied. The width of the resonance peaks is estimated as $x_w\sqrt{\gamma/\delta E_s}\ll\lambda_F$.

Rearranging Eq.~\eqref{gapend}, we can express the tunneling energy $\gamma$ in terms of the experimentally observable quantities $E_g$ and $\Delta$ (similarly to what is done in Refs.~\cite{vanHeck:2016,Reeg:2017_2} for the case of a bulk superconductor). However, due to the presence of the quantities $k_Fx_w$ and $k_Fd$, which would be impossible to determine experimentally, we can obtain only an order of magnitude estimate for $\gamma$ for the case of a finite superconductor. Assuming that $\sin^2(k_Fd)\sim1$, we find
\begin{equation} \label{gammaend}
\gamma_\text{edge}\sim\frac{E_g}{\Delta-E_g}\frac{\delta E_s}{(k_Fx_w)^2}.
\end{equation}
From Eq.~\eqref{gammaend}, it is clear that the lower bound on the tunneling strength needed to induce a sizable gap in the system (such that $E_g\sim\Delta$) is given by the level spacing $\delta E_s$.

\begin{figure}[t!]
\centering
\includegraphics[width=\linewidth]{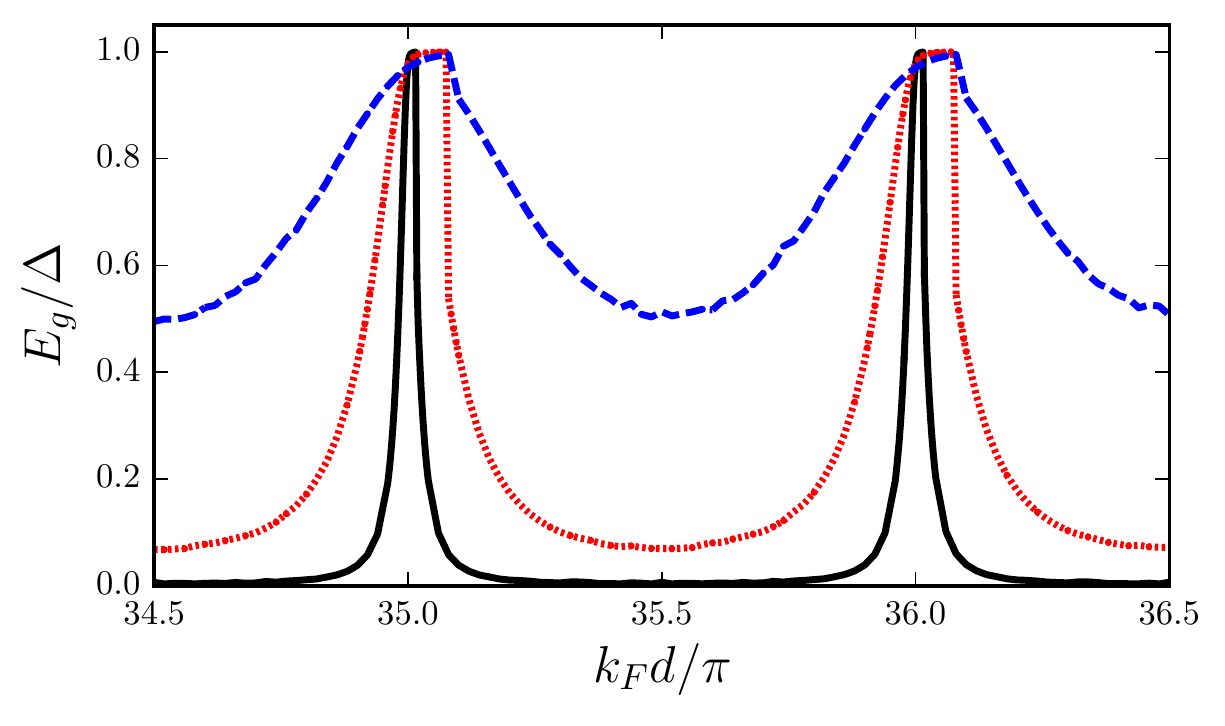}
\caption{\label{endanalytics} Excitation gap $E_g$ as a function of superconductor width $d$ for $\gamma=4\Delta$ (black), $\gamma=75\Delta$ (red), and $\gamma=1000\Delta$ (blue) when the nanowire is located near the edge of the superconductor. For weak tunneling strengths $\gamma\ll\delta E_s/(k_Fx_w)^2$, a small gap is induced for general $d$, with sharp (on the scale of the Fermi wavelength) resonance peaks near $k_Fd=\pi n$. A sizable gap is induced for all $d$ when $\gamma\gtrsim\delta E_s/(k_Fx_w)^2$. Remaining parameters chosen to be $k_F\xi_s=2\times10^4$, $m_w/m_s=0.02$, and $k_Fx_w=0.3$ [$\delta E_s/(k_Fx_w)^2\sim1000\Delta$].}
\end{figure}

\subsubsection{Numerical calculation of spectrum} \label{Sec3a2}

While we were able to solve for the excitation spectrum at energies $E<\Delta$ to determine the gap in certain limits [see Eq.~\eqref{specend}], it is much more difficult to solve for the full spectrum. Because the full spectrum $E(k_y)$ obeys a transcendental equation that cannot be solved analytically in general, we must resort to solving Eq.~\eqref{solvability} numerically.

In Fig.~\ref{endanalytics}, we plot the excitation gap $E_g$ as a function of superconductor width $d$. We calculate the gap numerically by computing the spectrum and finding the minimum of the lowest subband, allowing us to treat values of $k_Fd$ for which Eq.~\eqref{gapend} breaks down [namely, for $\sin(k_Fd)\to0$ and $\mu_\text{eff}<0$]. Overall, we find very good agreement between the numerical solution for the gap and the analytical form given in Eq.~\eqref{gapend}. For weak tunneling [Fig.~\ref{endanalytics}(a)], the gap is in general very small with very sharp resonance peaks around $k_Fd=\pi n$. As the tunneling is increased, the resonance peak is broadened and the size of the gap is generally shifted to larger values [Fig.~\ref{endanalytics}(b)]. When $\gamma\sim\delta E_s/(k_Fx_w)^2$, the gap is always of the same order as that of the superconductor, $E_g\sim\Delta$ [Fig.~\ref{endanalytics}(c)].

\begin{figure}[t!]
\centering
\includegraphics[width=\linewidth]{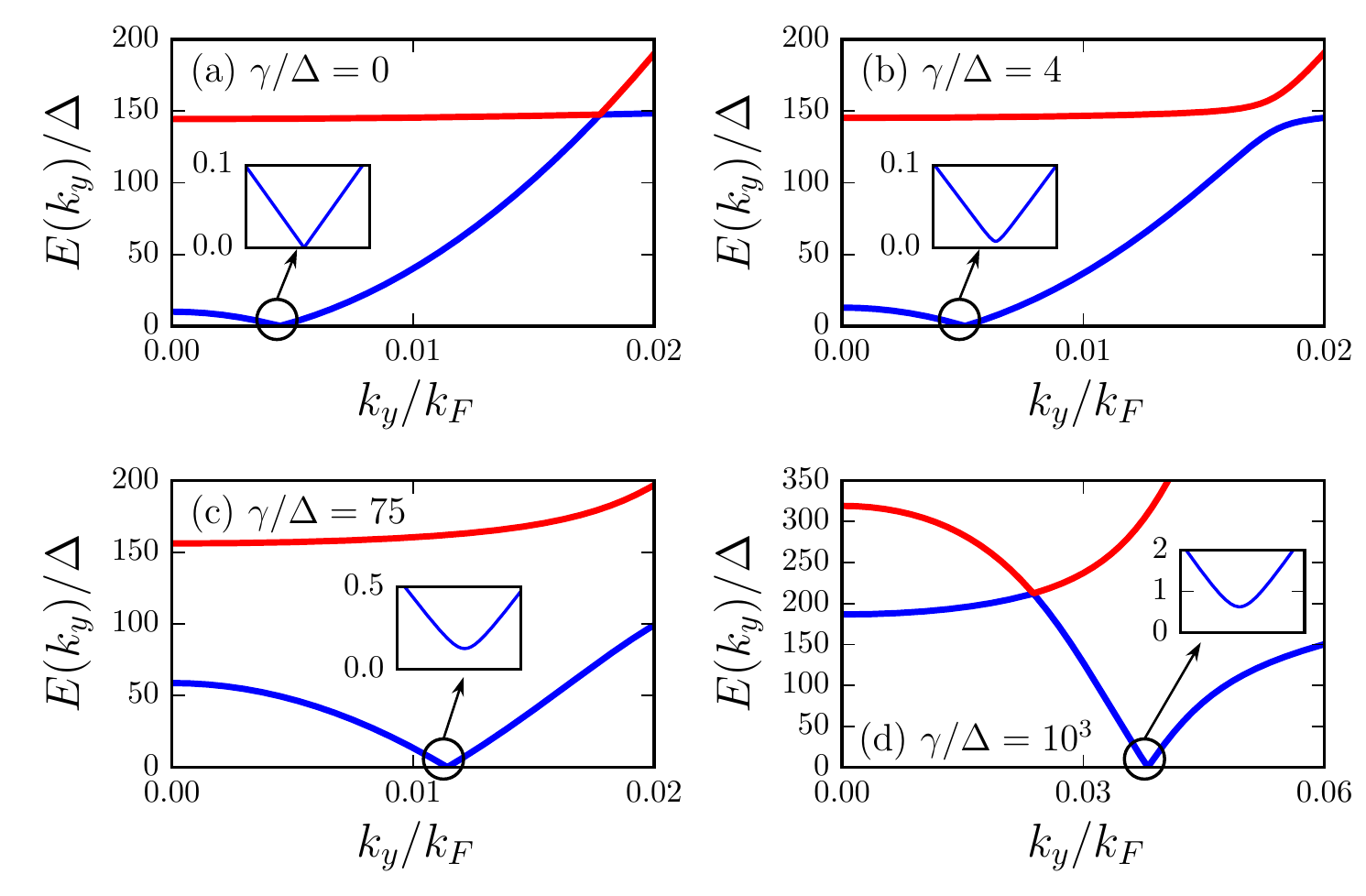}
\caption{\label{spectra} Lowest two subbands of excitation spectrum $E(k_y)$ for fixed $k_Fd=34.75\pi$ (away from resonance peaks of Fig.~\ref{endanalytics}) and tunneling strengths (a) $\gamma=0$, (b) $\gamma=4\Delta$, (c) $\gamma=75\Delta$, and (d) $\gamma=1000\Delta$. As the tunneling strength is increased, the nanowire band can more efficiently couple to the superconductor and the excitation gap is increased. The effective chemical potential of the wire $\mu_\text{eff}$ also increases with tunneling strength [see Eq.~\eqref{effparamsend}]; when $\mu_\text{eff}\gtrsim\delta E_s$, the nanowire and superconducting bands overlap and a large gap is induced [see panel (d)]. The remaining parameters are the same as in Fig.~\ref{endanalytics}.}
\end{figure}

To better understand the behavior of the gap as a function of $\gamma$, we plot the spectrum for various $\gamma$ and fixed superconductor width (chosen to be off resonance) in Fig.~\ref{spectra}. In the absence of tunneling [Fig.~\ref{spectra}(a)], there is a large separation in energy between the band of the nanowire and the lowest subband of the superconductor (a consequence of the fact that $\delta E_s\gg\mu_w$). As the tunneling strength is increased [Figs.~\ref{spectra}(b) and \ref{spectra}(c)], the effective chemical potential of the nanowire $\mu_\text{eff}$ increases and the two lowest subbands move closer in energy; as a result, the nanowire can more efficiently couple to the superconductor and the proximity-induced gap increases. When $\gamma\sim\delta E_s/(k_Fx_w)^2$ [Fig.~\ref{spectra}(d)], the tunneling is strong enough to overcome the large subband spacing of the superconductor. This creates significant overlap between the two lowest subbands of the system and a sizable excitation gap $E_g\sim\Delta$.

\subsubsection{Simple two-band model} \label{Sec3a3}

In this section, we present a simple two-band model which can be used to better understand the ``weak tunneling" limit $\gamma\ll\delta E_s/(k_Fx_w)^2$. In this limit, we can safely assume that the nanowire couples only to the lowest subband of the superconductor. Taking into account only the lowest superconducting subband, we can write down a simple tunneling Hamiltonian
\begin{equation}
H=\frac{1}{2}\int\frac{dk_y}{2\pi}\,\Psi^\dagger\begin{pmatrix}
\xi_k & 0 & -t & 0 \\
0 & -\xi_k & 0 & t \\
-t & 0 & \xi_n & \Delta \\
0 & t & \Delta & -\xi_n \\
\end{pmatrix}
\Psi,
\end{equation}
where $\Psi=(\Psi_w,\Psi_n)^T$ ($\Psi_w$ describes states in the nanowire, while $\Psi_n$ describes states in subband $n$ of the superconductor) and $t$ is a coupling between the two bands with dimensions of energy [note that this is not the same $t$ which was introduced in Eq.~\eqref{Ht}]. Quantization of the superconducting bands is accounted for through $\xi_n=k_y^2/2m_s-\mu_n$, with $\mu_n=\delta E_s(k_Fd-\pi n)$ for $n\in\mathbb{Z}^+$ [see Eq.~\eqref{barespec2}].

\begin{figure}[t!]
\centering
\includegraphics[width=\linewidth]{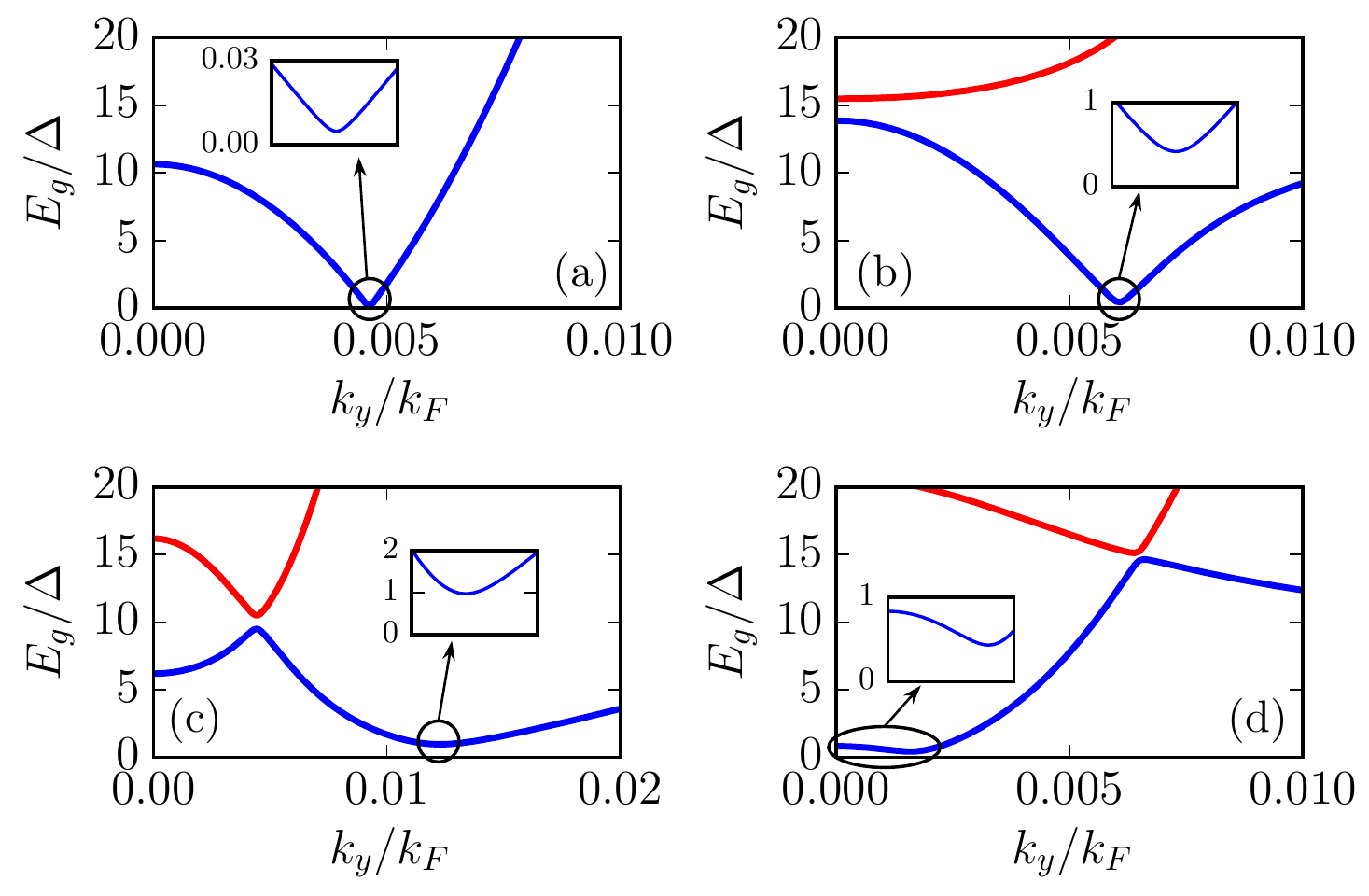}
\caption{\label{spectra2band} Spectrum of two-band model [Eq.~\eqref{effspec}] for fixed tunneling strength $t=10\Delta$ and different superconductor widths (a) $k_Fd/\pi=34.75$, (b) $k_Fd/\pi=34.98$, (c) $k_Fd/\pi=35$, and (d) $k_Fd/\pi=35.02$. For these choices of $k_Fd$, the relevant superconducting subband corresponds to $n=35$. The induced gap is sharply peaked around $k_Fd/\pi=35$, as even a small shift away from resonance leads to a drastic reduction in the size of the gap. This picture is consistent with the resonance behavior observed for weak tunneling in Fig.~\ref{endanalytics}. The remaining parameters are chosen as in Fig.~\ref{endanalytics}.}
\end{figure}

The corresponding spectrum is given by
\begin{equation} \label{effspec}
\begin{aligned}
2 E^2 &= \Delta^2 + \xi_k^2+  \xi_n ^2 +2t^2 \\
& \pm \sqrt{(\Delta^2 - \xi_k^2+\xi_n^2)^2+4t^2[\Delta^2+(\xi_k+\xi_n)^2]}.
\end{aligned}
\end{equation}
In general, $|\mu_n|\gg\mu_w,\Delta,t$ and we can expand Eq.~\eqref{effspec} to give
\begin{equation} \label{specoffresonance}
E=\pm\sqrt{\xi_k^2+t^4/\mu_n^2}.
\end{equation}
In this case, the lower subband takes on a superconducting dispersion with a small induced gap $E_g=t^2/\mu_n\ll\Delta$. The gap can only be enhanced when $|\mu_n|\lesssim\mu_w,\Delta,t$, which occurs only when a new superconducting subband becomes occupied, $k_Fd\approx \pi n$. While we cannot solve analytically for the gap in this limit, we find by plotting the spectrum that the gap is approximately $\Delta$.

We plot the spectrum Eq.~\eqref{effspec} for different superconductor widths $d$ in Fig.~\ref{spectra2band}. Away from resonance [Fig.~\ref{spectra2band}(a)], the lowest subband has a superconducting dispersion with very small induced gap [Eq.~\eqref{specoffresonance}]. As the resonance is approached [Fig.~\ref{spectra2band}(b)], the lowest superconducting subband becomes available to more strongly couple to the nanowire band and the gap is enhanced. On resonance [Fig.~\ref{spectra2band}(c)], overlap between the two subbands is maximal and the full gap $\Delta$ is opened. As a new subband in the superconductor becomes occupied and moves away from the nanowire band [Fig.~\ref{spectra2band}(d)], the gap is again suppressed by the large subband spacing in the superconductor. By plotting the spectrum, we indeed see that the excitation gap is sharply peaked as a function of $d$ around $k_Fd=\pi n$, consistent with the resonance behavior discussed in Sec.~\ref{Sec3}\ref{Sec3a1} and shown in Fig.~\ref{endanalytics}.

\subsection{Wire in middle of superconductor} \label{Sec3b}

If the wire is placed in the middle of the superconductor, $x_w=d/2$, the effective parameters of Eq.~\eqref{effparams} for energies $E<\Delta$ can be simplified to
\begin{equation} \label{effparamsmiddle}
\begin{gathered}
\Gamma=\left(1+\frac{\gamma}{\delta E_s}\frac{1}{2\cos^2(k_Fd/2)}\right)^{-1}, \\
\delta\mu=\gamma\tan(k_Fd/2).
\end{gathered}
\end{equation}
Again, we neglect the momentum dependence of the effective parameters by setting $\varphi=1$ and assume $|\sin(k_Fd)|\gg\Delta/\delta E_s$.

Since the parameters in Eq.~\eqref{effparamsmiddle} are independent of energy and momentum, the spectrum is again given by Eq.~\eqref{specend}. Assuming that $\mu_\text{eff}>0$, we find an excitation gap
\begin{equation} \label{gapmiddle}
E_g/\Delta=1-\left(1+\frac{\gamma}{\delta E_s}\frac{1}{2\cos^2(k_Fd/2)}\right)^{-1}.
\end{equation}
While the gap can still be small for $\gamma\gg\Delta$, there are two important differences when comparing to the case when the nanowire is at the edge of the superconductor [Eq.~\eqref{gapend}]. First, the gap is no longer suppressed by the factor $(k_Fx_w)^2\ll1$ which originated from the smallness of the superconducting wave function near the edge. Instead, without any fine-tuning of the superconductor width [$\cos^2(k_Fd/2)\sim1$], a sizable gap can be induced for $\gamma\gtrsim\delta E_s$. Second, the periodicity of the gap as a function of $k_Fd$ is twice as large. In the weak tunneling limit $\gamma\ll\delta E_s$, resonance peaks are half as frequent and occur near $k_Fd=\pi(2n+1)$ ($n\in\mathbb{Z}^+$). The width of each resonance peak is larger than in the case of the nanowire at the edge of the superconductor, but it is still much smaller than the Fermi wavelength, $\lambda_F\sqrt{\gamma/\delta E_s}\ll\lambda_F$.

\begin{figure}[t!]
\includegraphics[width=\linewidth]{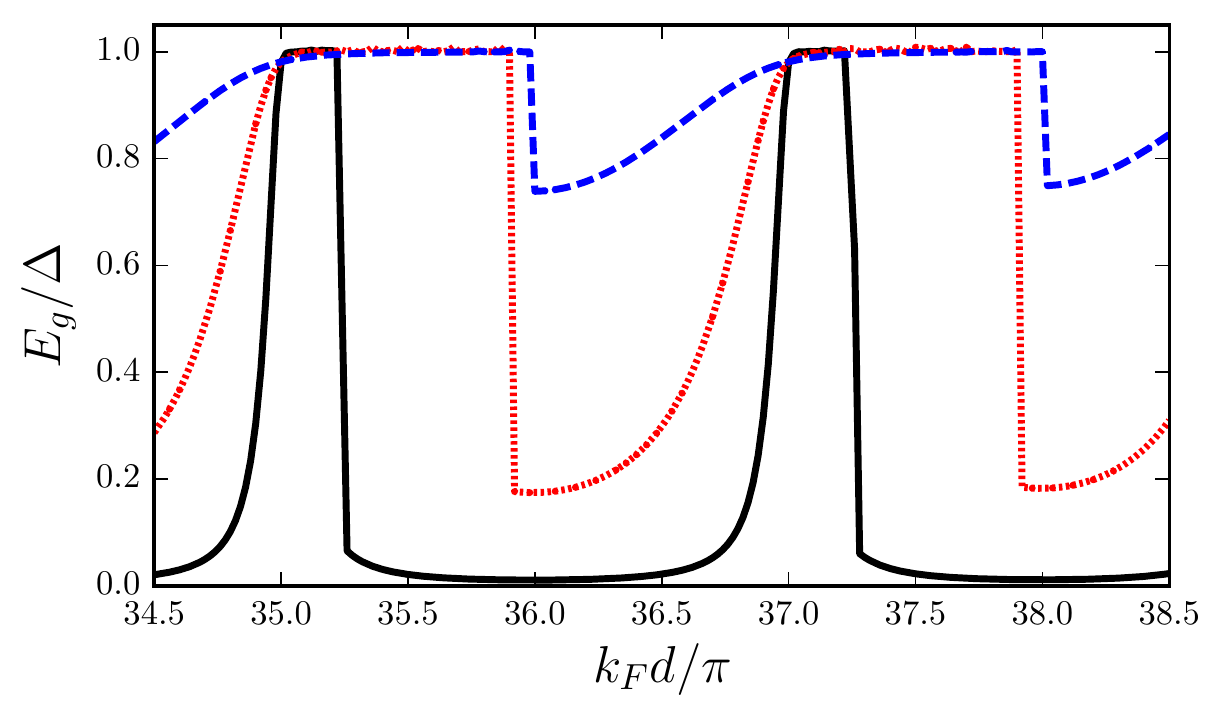}
\caption{\label{middleanalytics} Excitation gap $E_g$ as a function of superconductor width $d$ for $\gamma=4\Delta$ (black), $\gamma=75\Delta$ (red), and $\gamma=1000\Delta$ (blue) when the nanowire is located in the middle of the superconductor. For all tunneling strengths, we observe extended plateau regions as a function of $d$; these plateaus correspond to instances when the nanowire is completely depleted ($\mu_\text{eff}<0$) and the excitation gap is therefore determined by the gap on the lowest superconducting subband. Away from the plateau regions, a sizable gap is induced in the nanowire for all $d$ when $\gamma\gtrsim\delta E_s$. We also find that the periodicity of the gap as a function of $d$ is increased by a factor of two compared to when the nanowire is located at the edge of the superconductor [Fig.~\ref{endanalytics}]. The remaining parameters are chosen as in Fig.~\ref{endanalytics}, corresponding to $\delta E_s\sim100\Delta$.}
\end{figure}

The excitation gap $E_g$ is plotted as a function of $d$ for several different tunneling strengths $\gamma$ in Fig.~\ref{middleanalytics}. Similarly to Sec.~\ref{Sec3}\ref{Sec3a2}, we solve for the gap numerically by computing the full spectrum $E(k_y)$ and finding the minimum. Again, all parameters are chosen the same as in Fig.~\ref{endanalytics} (except for $x_w=d/2$). As previously discussed, the periodicity of the gap as a function of $k_Fd$ is increased by a factor of two compared with the case when the nanowire is located at the edge of the superconductor. We see that the gap is maximized near $k_Fd/\pi=2n+1$ and is minimized near $k_Fd/\pi=2n$. This result can be inferred from the structure of the wave function corresponding to the lowest quantized subband of the superconductor in the absence of the nanowire. For $k_Fd/\pi=2n+1$, there is a superconducting subband available at low energies with which the nanowire can couple; because the wave function of this subband is extremal at $x=d/2$, the nanowire efficiently couples and a large gap is induced. For $k_Fd/\pi=2n$ there is again a superconducting subband at low energies; however, the wave function of this subband has a node at $x=d/2$ and does not couple efficiently to the nanowire. In this case, a sufficiently large gap is opened only when the tunneling is very strong.

\begin{figure}[t!]
\includegraphics[width=\linewidth]{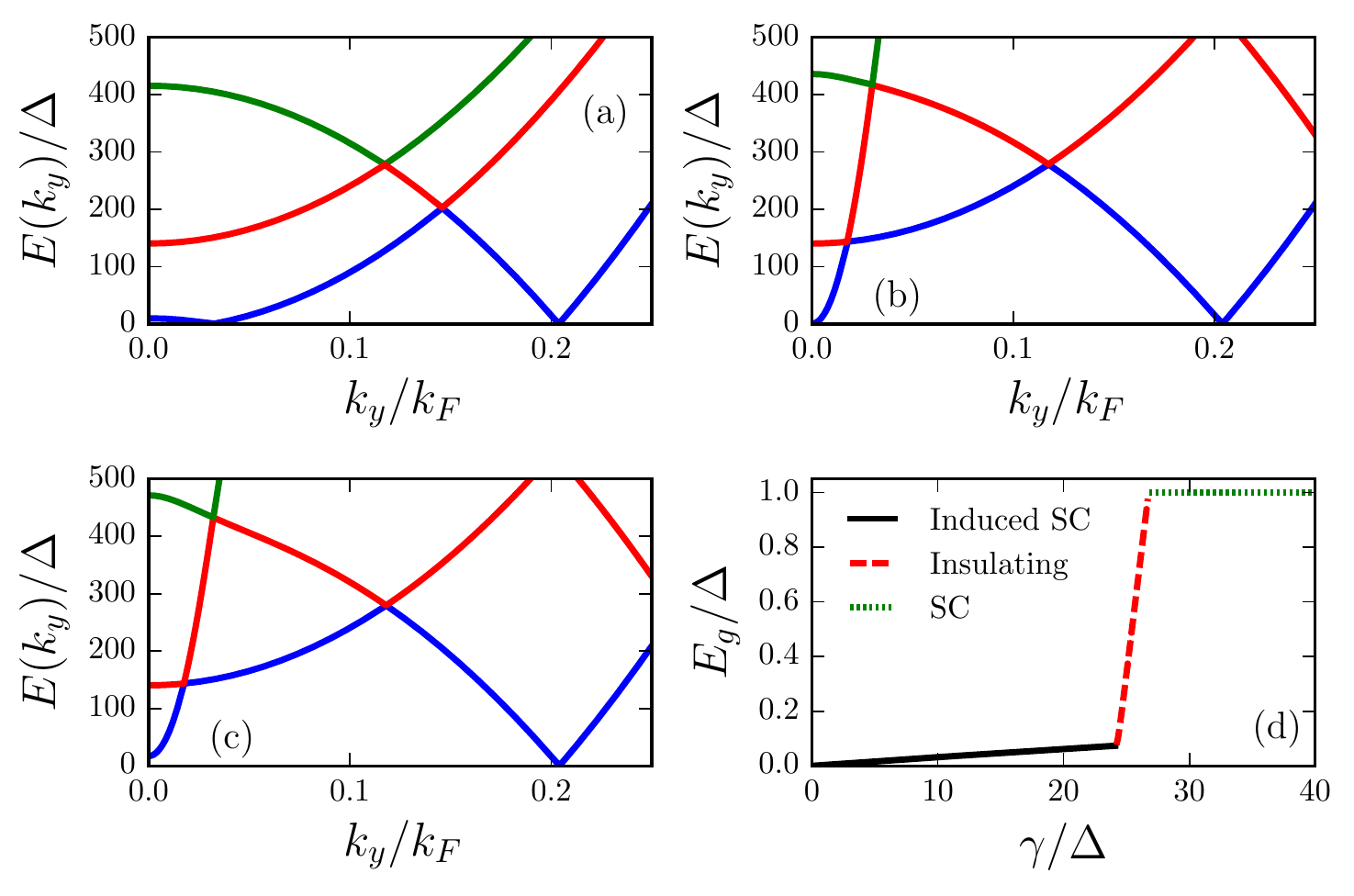}
\caption{\label{spectramiddle} Excitation spectrum for $k_Fd/\pi=35.75$ (corresponding to one of the plateau regions of Fig.~\ref{middleanalytics}) with (a) $\gamma=0$, (b) $\gamma=25\Delta$, and (c) $\gamma=75\Delta$. As the tunneling strength is increased, the nanowire band becomes depleted and eventually becomes insulating. (d) Excitation gap as a function of tunneling strength $\gamma$. We also denote whether the minimum gap in the spectrum is an induced superconducting gap in the nanowire band at finite $k_y$ [see (a)], an insulating gap in the nanowire band at $k_y=0$ [see (b)], or the gap $\Delta$ on the lowest occupied subband of the superconductor [see (c)]. The remaining parameters are chosen as in Fig.~\ref{endanalytics}.}
\end{figure}

We also find that as the tunneling strength is increased, extended plateaus emerge as a function of $k_Fd$. To better understand this behavior, in Figs.~\ref{spectramiddle}(a)--\ref{spectramiddle}(c) we plot the spectrum for different $\gamma$ choosing $k_Fd/\pi=35.75$. As the tunneling strength is increased from $\gamma=0$ [Fig.~\ref{spectramiddle}(a)], a superconducting gap is induced on the band which originates in the nanowire. However, at the same time, the nanowire band gets depleted. For some critical tunneling strength, the nanowire band becomes depleted completely and enters an insulating phase. The minimum excitation gap is then given by the insulating gap in the nanowire at $k_y=0$ [Fig.~\ref{spectramiddle}(b)]. As the tunneling strength is increased further and the insulating gap on the nanowire band exceeds $\Delta$, the minimum excitation gap is determined by the lowest occupied subband of the superconductor [Fig.~\ref{spectramiddle}(c)]. This behavior can be understood as follows. The depletion of the nanowire band can be inferred from $\delta\mu$ given in Eq.~\eqref{effparamsmiddle}. When the wire is in the middle of the superconductor, $\delta\mu<0$ for precisely half of a period, including for $k_Fd/\pi=35.75$. In the strong tunneling limit, $|\delta\mu|\gg\mu_n$ in general and $\mu_\text{eff}<0$ (i.e., the nanowire becomes insulating) also for half of a period (and, as shown in Fig.~\ref{middleanalytics}, the plateau extends over half of a period in the strong tunneling limit). We also find that the lowest subband of the superconductor (corresponding to $n=36$) remains almost completely unaffected by tunneling. This is consistent with our previous analysis and results from the fact that this subband cannot efficiently couple to the nanowire because the corresponding wave function has a node at $x=d/2$. Rather, the nanowire band couples most efficiently to the second-lowest subband (corresponding to $n=35$). While the presence of the nanowire modifies this subband somewhat in the vicinity of $k_y=0$, it has no effect on the gap of this subband at finite $k_y$. These findings are summarized in Fig.~\ref{spectramiddle}(d), where we plot the excitation gap as a function of tunneling strength for $k_Fd/\pi=35.75$.

Using our result for the gap [Eq.~\eqref{gapmiddle}], we provide an order of magnitude estimate for the tunneling strength $\gamma$ assuming that the superconductor width $d$ is tuned outside of the plateau region and satisfies $\cos^2(k_Fd/2)\sim1$. In this case, we estimate
\begin{equation} \label{gammamiddle}
\gamma_\text{middle}\sim\frac{E_g}{\Delta-E_g}\delta E_s.
\end{equation}
Compared with Eq.~\eqref{gammaend}, we find that a sizable gap ($E_g\sim\Delta$) can be induced with a much smaller tunneling strength when the wire is placed in the middle of the superconductor.

\begin{figure}[t!]
\includegraphics[width=\linewidth]{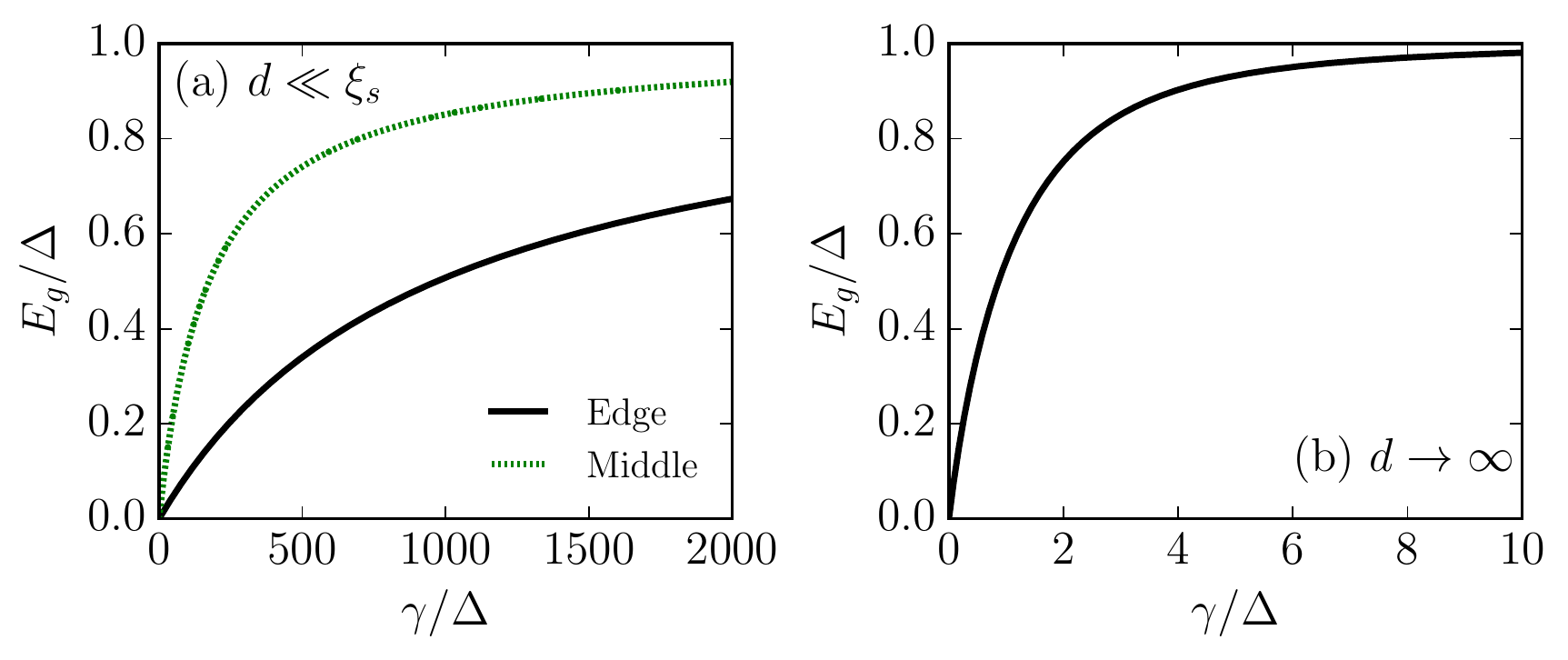}
\caption{\label{Egvsgamma} Excitation gap $E_g$ as a function of tunneling strength $\gamma$ for fixed $d$. (a) Width of superconductor much smaller than its coherence length, $k_Fd/\pi=36.5$ ($d/\xi_s=5.7\times10^{-3}$), for both the case when the nanowire is located near the edge of the superconductor [Eq.~\eqref{gapend}, taking $k_Fx_w=0.3$] and in the middle of the superconductor [Eq.~\eqref{gapmiddle}]. (b) Bulk superconductor, $d\to\infty$ [Eq.~\eqref{gammabulk}]. In order to induce a sizable gap in the nanowire, the tunneling strength needed when $d\ll\xi_s$ is orders of magnitude larger than the bulk case.}
\end{figure}

Finally, we compare three different cases by plotting the induced gap in the nanowire as a function of $\gamma$ in Fig.~\ref{Egvsgamma}. In Fig.~\ref{Egvsgamma}(a), we show a direct comparison between the case of a nanowire located at the edge of the superconductor and the case of a nanowire located in the middle. We also contrast our result for a finite superconductor against that for a bulk superconductor by plotting the induced gap as a function of $\gamma$ for the latter case in Fig.~\ref{Egvsgamma}(b). For a bulk system, the induced gap obeys the equation \cite{vanHeck:2016,Reeg:2017_2}
\begin{equation} \label{gammabulk}
\gamma_\text{bulk}=E_g\sqrt{\frac{\Delta+E_g}{\Delta-E_g}}.
\end{equation}
We see that to open a sizable gap in a finite superconductor, a tunneling strength which is several orders of magnitude larger than in the bulk case is needed.

\section{Tight-binding model} \label{Sec4}

In this section, we check our analytical results by comparing them with a numerical tight-binding model for proximity-induced superconductivity \cite{Rainis:2013,Klinovaja:2015,Yang:2016} in the geometry shown in Fig.~\ref{setup}. The system is again assumed infinite in the $y$-direction, so that the Hamiltonian takes a block-diagonal form in momentum $k_y$, $H=\sum_{k_y} H_{k_y}$.  The size of the superconductor in the $x$ direction is $N a_x$ ($a_{x,y}$ are lattice constants), while the size of the nanowire is taken to be a single site. The Hamiltonian of the superconductor is given by
\begin{equation}
\begin{aligned}
H_{s,k_y}&= \sum_\sigma \sum_{i=1}^{N}\biggl\{ [\mu_s- 2 t_0 \cos (k_y a_y)] c_{k_y,i,\sigma}^\dagger c_{k_y,i,\sigma} \\
&-(t_0c_{k_y,i+1,\sigma}^\dagger c_{k_y,i,\sigma} -\Delta c_{k_y ,i ,\uparrow}^\dagger c_{-k_y ,i ,\downarrow}^\dagger + H.c.)\biggr\},
\end{aligned}
\end{equation}
where $c_{k_y,i,\sigma}$ destroys a state of momentum $k_y$ and spin $\sigma$ in the superconductor at site $i$, $t_0$ is the hopping amplitude, $\mu_s$ is the chemical potential (calculated from the bottom of the band), and $\Delta$ is the pairing potential. The Hamiltonian of the nanowire is given by
\begin{equation}
H_{w,k_y}=\sum_\sigma[\mu_w-2t_w\cos(k_ya_y)]b^\dagger_{k_y,\sigma}b_{k_y,\sigma},
\end{equation}
where $b_{k_y,\sigma}$ destroys a state of momentum $k_y$ and spin $\sigma$ in the nanowire, $\mu_w$ is the chemical potential, and $t_w$ is the hopping amplitude. The nanowire is coupled to the superconductor at site $j$,
\begin{equation}
H_{t,k_y}=-t\sum_\sigma(c^\dagger_{k_y,j,\sigma}b_{k_y,\sigma}+H.c.),
\end{equation}
where $t$ is the tunneling amplitude between the nanowire and superconductor. We assume that tunneling preserves both spin and momentum.

\begin{figure}[t!]
\centering
\includegraphics[width=\linewidth]{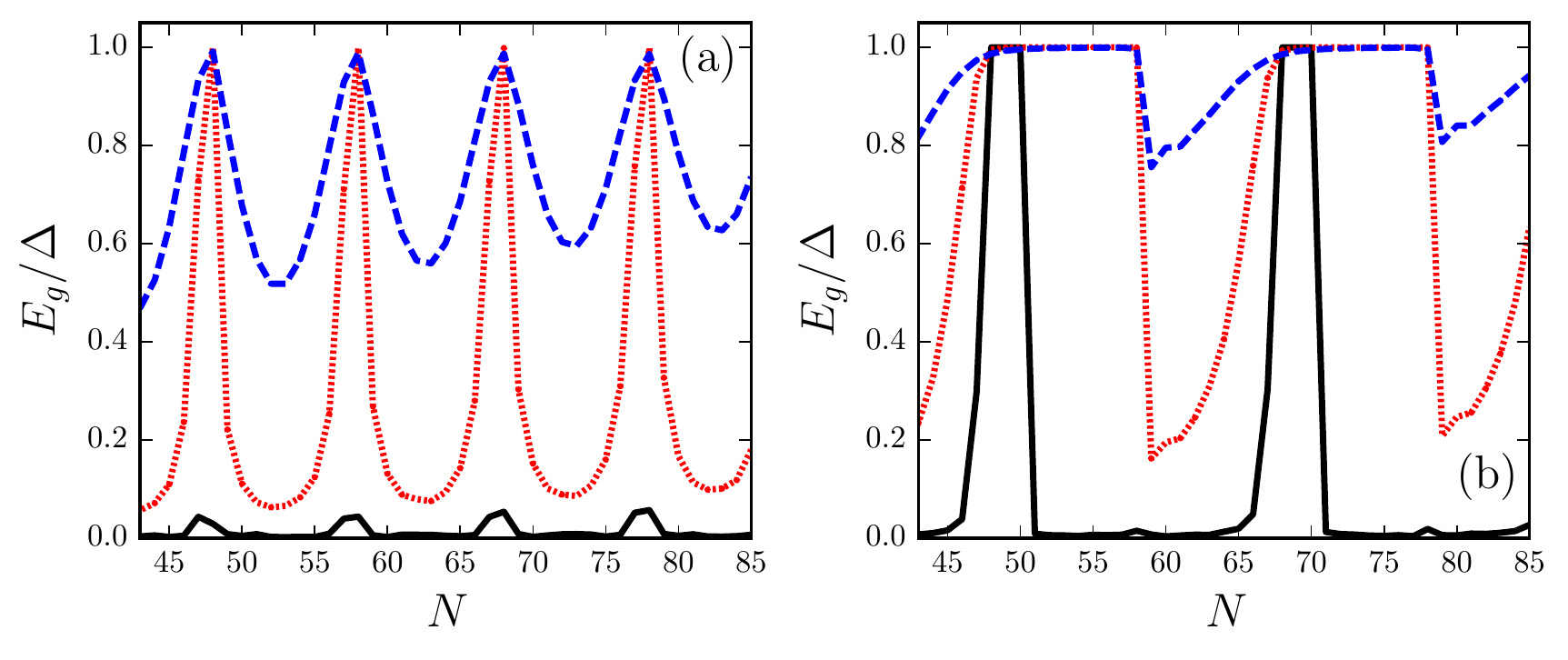}
\caption{\label{numerics} Excitation gap $E_g$ as a function of superconductor width $N$, calculated by numerically diagonalizing a tight-binding Hamiltonian for tunneling strengths $t=0.006t_0$ (black), $t=0.05t_0$ (red), and $t=0.2t_0$ (blue). (a) Nanowire at edge of superconductor. Similarly to Fig.~\ref{endanalytics}, we observe sharp resonance peaks in the weak tunneling limit. Note that for the black curve, the width of the resonance peak is narrower than a single site. (b) Nanowire in middle of superconductor. Similarly to Fig.~\ref{middleanalytics}, we observe both extended plateau regions and a doubling of the periodicity of the gap as a function of superconductor width. In both plots, we choose $\mu_s=0.1t_0$, $\mu_w=10^{-4}t_0$, $\Delta=10^{-5}t_0$, and $t_w=50t_0$ (all parameters are chosen to be consistent with those in Fig.~\ref{endanalytics}).}
\end{figure}

We consider two separate cases. First, the nanowire is placed at the end of the superconducting chain ($j=1$). Whereas analytically we were unable to place the nanowire strictly at the edge of the superconductor due to its vanishing width, we can do so in the tight-binding formulation (the nanowire has a finite width of one site). Second, the nanowire is placed in the middle of the superconductor ($j=N/2$). The results of our tight-binding calculation are shown in Fig.~\ref{numerics}. We plot the excitation gap as a function of superconductor width choosing $\mu_s=0.1t_0$, $\mu_w=10^{-4}t_0$, $\Delta=10^{-5}t_0$, and $t_w=50t_0$ (all parameters are chosen to coincide with those used previously in the analytical calculation). For Fig.~\ref{numerics}(a), which corresponds to $j=1$, we find very good agreement with the analytics of Sec.~\ref{Sec3a} and Fig.~\ref{endanalytics}. We note that the resonance peaks in the curve corresponding to weak tunneling [black curve in Fig.~\ref{numerics}(a)] are narrower than a single site and therefore do not appear in the numerics. For Fig.~\ref{numerics}(b), which corresponds to $j=N/2$, we find very good agreement with the analytics of Sec.~\ref{Sec3b} and Fig.~\ref{middleanalytics}. Notably, a sizable gap $E_g\sim\Delta$ is seen only when the tunneling strength $t$ exceeds the chemical potential of the superconductor.

\section{Topological Criterion} \label{Sec5}
To access the topological phase, we now assume that the nanowire has Rashba spin-orbit coupling and a Zeeman splitting that results from the application of an external magnetic field $B$ parallel to the nanowire, corresponding to the typical setup for realizing topological superconductivity \cite{Oreg:2010,Lutchyn:2010}. The Hamiltonian of the nanowire in this case is given by 
\begin{equation}
H_w=\frac{1}{2}\int\frac{dk_y}{2\pi}\,\Psi^\dagger(k_y){\mathcal{H}}_w(k_y)\Psi(k_y),
\end{equation}
where $\Psi(k_y)=[\psi_\uparrow(k_y),\psi_\downarrow(k_y),\psi_\uparrow^\dagger(-k_y),\psi_\downarrow^\dagger(-k_y)]^T$. The Hamiltonian density, which is a $4\times4$ matrix in Nambu $\otimes$ spin space, is given by
\begin{equation} \label{SOCfield}
{\mathcal{H}}_w(k_y)=\xi_k\tau_z-\alpha k_y\sigma_z-\Delta_Z\tau_z\sigma_x,
\end{equation}
where $\alpha$ is the Rashba constant, $\Delta_Z=g\mu_BB/2$ is the Zeeman splitting ($g$ is the $g$-factor of the nanowire and $\mu_B$ is the Bohr magneton), and $\sigma_{x,y,z}$ are Pauli matrices acting in spin space. Following convention, we neglect the effect of the external magnetic field on the superconductor. Generalization of the solution of the BdG equation given in Eq.~\eqref{WF} to the case where one additionally has to account for the spin degree of freedom is straightforward. The addition of spin-orbit coupling and Zeeman splitting simply modifies the retarded Green's function which enters the boundary conditions [Eqs.~\eqref{BCs}], ${G}_0^R=(E-{\mathcal{H}}_w+i0^+)^{-1}$. Solving the boundary conditions, we find that the self-energy given in Eq.~\eqref{selfenergy} still holds, with the simple replacement $\Delta\tau_x\to-\Delta\tau_y\sigma_y$ to account for the spin-singlet nature of the induced pairing \cite{Schrade:2017}.

Given the self-energy, we find that the excitation spectrum is determined from the implicit equation
\begin{align} \label{spectrumtop}
&E^2/\Gamma^2=\Delta_Z^2+\Delta^2\left(1/\Gamma-1\right)^2+(\xi_k-\delta\mu)^2+\alpha^2k_y^2 \nonumber\\
	&\pm2\sqrt{\Delta_Z^2\Delta^2\left(1/\Gamma-1\right)^2+(\xi_k-\delta\mu)^2(\Delta_Z^2+\alpha^2k_y^2)}.
\end{align}
The critical Zeeman splitting needed to close the excitation gap at $k_y=0$, which we find by setting $k_y=E=0$ in Eq.~\eqref{spectrumtop}, is found to be
\begin{equation} \label{topcriterion}
\begin{aligned}
\Gamma\Delta_Z^c&=\sqrt{\Gamma^2(\mu_w+\delta\mu)^2+E_g^2},
\end{aligned}
\end{equation}
where we replace $\Delta(1-\Gamma)=E_g$, noting that this replacement is strictly valid only when $\delta E_s\gg\Delta$. [The explicit forms of $\delta\mu(0,0)$ and $\Gamma(0,0)$ are given in Eq.~\eqref{effparams}.]
If the chemical potential shift $\delta\mu$ can be compensated by tuning the chemical potential of the wire (such that $\mu_w+\delta\mu=0$), then the critical field strength is $\Delta_Z^c\sim E_g/\Gamma$ (this is a similar criterion used, for example, to analyze the data of Ref.~{\cite{Deng:2016}}; note that the critical Zeeman splitting needed is larger than the induced gap $E_g$). If, however, the chemical potential shift is made too large ($|\delta\mu|\gg\mu_w$), the critical Zeeman splitting is determined solely by this shift, $\Delta_Z^c\sim|\delta\mu|$, and the finite size of the superconductor pushes the topological threshold to significantly higher magnetic field strength. We will provide numerical estimates in the next section to argue that the latter case is more relevant to thin superconducting shells that have a large level spacing $\delta E_s\gg\Delta$.

\section{Relation to experiments with epitaxial superconducting shells} \label{Sec6}

\begin{figure}[t!]
\centering
\includegraphics[width=.4\linewidth]{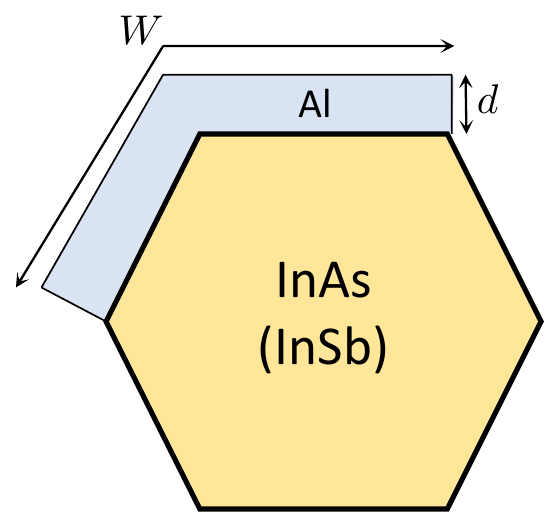}
\caption{\label{experiment} Cross-section of a hexagonal nanowire (InAs or InSb) coupled to an epitaxially grown thin superconducting shell (Al), similar to the devices studied in Refs.~\cite{Albrecht:2016,Deng:2016,Gazibegovic:2017}. The superconducting shell has thickness $d\sim10$ nm and width $W\sim100$ nm.}
\end{figure}

In this section, we argue that the theoretical model that we have considered to this point is applicable to recent experiments studying InAs or InSb nanowires strongly coupled to thin superconducting Al shells (see Fig.~\ref{experiment}) \cite{Chang:2015,Albrecht:2016,Deng:2016,Gazibegovic:2017}. We also provide an order of magnitude estimate for the level spacing of the shell, which determines the critical field strength needed to reach the topological phase in such a setup.

First, in order to treat the nanowire as one-dimensional, we assume that the wave function is uniform across the entire cross-section of the nanowire (which spans roughly $\sim100$ nm). Second, because we have replaced the nanowire cross-section by a single point, we must also neglect the width of the superconducting shell, $W\sim100$ nm (see Fig.~\ref{experiment}). To justify this assumption, let us define a phenomenological tunneling strength that originates from the coupling between the nanowire and superconductor along this dimension [call it $t(W)$]. Such a tunneling should be given by the product of the nanowire and superconducting wave functions, integrated over the width,
\begin{equation}
t(W)\sim\int_0^Wdz\,\psi_w^*(z)\psi_s(z).
\end{equation}
Here, we denote the dimension along the width of the shell by $z$, imagining for simplicity that the shell cross-section is rectangular rather than kinked. The superconducting wave function is quantized as $\psi_s\sim\sin(k_sz)/\sqrt{W}$, where $k_s=n\pi/W$ and $1/\sqrt{W}$ is a normalization factor. The wave function of the nanowire is uniform, but it also must be normalized, $\psi_w\sim1/\sqrt{W_w}$. Because the normalization factor of the nanowire wave function should scale with the width of the shell, $W_w\sim W$, the effective tunneling strength is given by
\begin{equation}
t(W)\sim\frac{1}{W}\int_0^Wdz\,\sin(k_sz)\sim1.
\end{equation}
Because the tunneling does not scale with $W$, this additional dimension is unimportant and can be neglected.

We now turn our attention toward applying our results to make qualitative predictions about the experimental setup. The relevant geometrical parameter (which corresponds to $d$) is the thickness of the superconducting shell, $d\sim10$ nm. For shells of this thickness, we estimate a level spacing $\delta E_s=\hbar v_F/d\sim10$ meV, where we take $v_F\sim10^6$ m/s for Al.

Given the level spacing of the shell, we can also estimate the tunneling strength needed to induce a gap $E_g\sim\Delta$. Because the experimental systems universally exhibit sizable induced gaps (and therefore it is safe to assume that the shell thicknesses are not fine-tuned to a resonance point), we use Eq.~\eqref{gammaend} to estimate
\begin{equation}
\gamma_\text{edge}\gtrsim100\text{\,meV},
\end{equation}
where we take $(k_Fx_w)^2\sim0.1$.

Therefore, the chemical potential shift induced by tunneling is estimated from Eq.~\eqref{effparamsend} as
\begin{equation}
|\delta\mu(0,0)|\sim\gamma(k_Fx_w)\gtrsim30\text{\,meV}.
\end{equation}
Given that the characteristic energy scale of the nanowire is the spin-orbit energy ($E_{so}=m_w\alpha^2/2\hbar^2$), which takes typical values $E_{so}\sim0.1-1$ meV in semiconducting nanowires \cite{vanWeperen:2015,Shabani:2016,Lutchyn:2017}, the topological phase transition is controlled entirely by tunneling, $\Delta_Z^c\sim|\delta\mu(0,0)|\gtrsim30$ meV. Such a large Zeeman energy corresponds to a critical field strength of $B_c\gtrsim60$ T for Al/InAs ($g_\text{InAs}\sim20$) and $B_c\gtrsim30$ T for Al/InSb ($g_\text{InSb}\sim40$). In either case, the magnetic field threshold needed to reach the topological phase greatly exceeds the field strength at which superconductivity in Al is destroyed (which occurs for $B\sim2$ T).

As discussed previously, in order to negate the effect of the tunneling-induced shift in the chemical potential, one must tune to $\mu_w=-\delta\mu(0,0)$. However, this is difficult in practice because $|\delta\mu(0,0)|\gg E_{so}$. If $\delta\mu(0,0)>0$, it is not possible to properly tune the chemical potential without completely depleting the semiconducting band; if $\delta\mu(0,0)<0$, the nanowire would have to be gated outside of the regime for which it is semiconducting.

We note that there are several aspects of the experimental setup which are not accounted for by our simple model. For example, our model does not include an additional renormalization of the nanowire $g$-factor by the superconductor beyond that which is contained in Eq.~{\eqref{topcriterion}}. This is because the $g$-factor in the superconductor was taken to be $g_s=0$ [and therefore no Zeeman term enters the self-energy of Eq.~{\eqref{selfenergy}}]. However, the additional renormalization should be small, as we estimate the effective $g$-factor of the wire as $\Gamma g+(1-\Gamma)g_s\approx\Gamma g$ given that $g\gg g_s$ and $\Gamma\sim(1-\Gamma)\sim1$ \cite{comment}. We also do not account for any orbital effects in the nanowire, as the nanowire was taken to have zero width. In the experimental setup, however, the nanowire has a diameter $W_w\sim100$ nm; compared with a typical cyclotron radius for electrons in the wire $r_c=m_wv_{Fw}/eB\sim10$ nm (evaluated for a field strength $B\sim1$ T and Fermi velocity $v_{Fw}\sim10^5$ m/s), orbital effects can be non-negligible \cite{Winkler:2017}. Additionally, we do not consider the superconductor to be disordered. It has been shown that disorder in a bulk superconductor can be detrimental to the proximity effect \cite{Cole:2016,Hui:2015}; however, this does not seem to be an issue experimentally, as sizable hard proximity-induced gaps are universally observed \cite{Chang:2015,Albrecht:2016,Deng:2016,Zhang:2017,Gazibegovic:2017} along with ballistic transport in the absence of a magnetic field \cite{Zhang:2017}. Relatedly, given the fact that applied magnetic fields are small, we do not determine $\Delta(x)$ self-consistently as done, for example, for Shiba states, where the exchange interaction (effective Zeeman field) is comparable to the Fermi energy \cite{Flatte:1997,Salkola:1997,Balatsky:2006,Meng:2014,Hoffman:2015}. Finally, we do not account for the fact that there may be multiple subbands in the nanowire which contribute to transport. While this possibility seems to be excluded by the fact that typical nanowires exhibit a quantized conductance of $2e^2/h$ in the ballistic transport limit over a wide range of gate voltages \cite{Kammhuber:2016,Zhang:2017}, it could be that with the introduction of the large tunneling energy scale, higher subbands can become important. The intersubband spacing in an InSb nanowire was estimated in the absence of a magnetic field and a superconducting shell to be $\sim20$ meV in Ref.~\cite{Kammhuber:2016}, so we cannot rule out the possibility that a $|\delta\mu(0,0)|\sim30$ meV simply places the chemical potential into a higher subband. However, this possibility requires a more in-depth theoretical treatment that cannot be captured by a single-band model.

\section{Conclusions} \label{Sec7}
We have studied the proximity effect in a semiconducting nanowire strongly coupled to a thin superconducting shell with thickness $d$. We have shown that finite-size effects become detrimental to the induced superconductivity when the level spacing of the shell $\delta E_s\sim\hbar v_F/d$ exceeds its gap, $\delta E_s\gg\Delta$. In this limit, a large tunneling strength $\gamma\gtrsim\delta E_s$ is needed to overcome the level spacing to induce a gap in the nanowire (we estimate $\gamma\gtrsim100$ meV for typical experimental setups with $d\sim10$ nm). In turn, this large tunneling energy induces a very large shift in the effective chemical potential of the nanowire ($\delta\mu\gtrsim30$ meV) which would be difficult to compensate by gating.

In order to overcome the detrimental finite-size effects, the thickness of the superconducting shell should be made larger than its coherence length, $d\gg\xi_s$. In this limit, the level spacing and chemical potential shift become negligible ($\delta E_s,\delta\mu\ll\Delta$), and a sizable proximity gap can be induced with a much smaller tunneling energy ($\gamma\sim\Delta$). However, this requirement may prove problematic when using Al, which has a very long coherence length $\xi_s\sim1$ $\mu$m, to induce superconductivity. It would therefore be beneficial to choose a superconductor with shorter coherence length (for example, Nb has $\xi_s\sim10$ nm).

\acknowledgments
This work was supported by the Swiss National Science Foundation and the NCCR QSIT. 

\newpage

\appendix
\section{Solvability Condition and Self-Energy} \label{appA}
In this appendix, we show how to arrive at the solvability condition given in Eq.~\eqref{solvability} of the main text. The boundary conditions that must be imposed on the wave function of Eq.~\eqref{WF} are given in Eq.~\eqref{BCs}. These boundary conditions can be significantly simplified by assuming that $\mu_s\gg\Delta,E$. In this limit, we can make a semiclassical approximation $k_\pm=(k_F^2-k_y^2\pm2m_si\Omega)^{1/2}\approx k_F\varphi\pm i\Omega/(v_F\varphi)$, where $\varphi=(1-k_y^2/k_F^2)^{1/2}$ ($0<\varphi\leq1$). The boundary conditions in Eq.~\eqref{BCs} can be written in matrix form as $Mc=0$, where $c=(c_1,c_2,c_3,c_4)^T$ and
\begin{widetext}
\begin{equation} \label{matrix}
\begin{aligned}
M&=\left(\begin{array}{cc}
	u_0\phi\cos(k_+x_w) & v_0\phi\cos(k_-x_w) \\
	v_0\phi\cos(k_+x_w) & u_0\phi\cos(k_-x_w) \\
	-u_0\sin(k_+x_w) & -v_0\sin(k_-x_w) \\
	-v_0\sin(k_+x_w) & -u_0\sin(k_-x_w)
	\end{array}\right. \cdots \\
	&\cdots\left.\begin{array}{cc}
		u_0\left(\varphi\cos[k_+(d-x_w)]+\frac{2\gamma}{E-\xi_k}\sin[k_+(d-x_w)]\right) & v_0\left(\varphi\cos[k_-(d-x_w)]+\frac{2\gamma}{E-\xi_k}\sin[k_-(d-x_w)]\right) \\
		v_0\left(\varphi\cos[k_+(d-x_w)]-\frac{2\gamma}{E+\xi_k}\sin[k_+(d-x_w)]\right) & u_0\left(\varphi\cos[k_-(d-x_w]-\frac{2\gamma}{E+\xi_k}\sin[k_-(d-x_w)]\right) \\
		u_0\sin[k_+(d-x_w)] & v_0\sin[k_-(d-x_w)] \\
		v_0\sin[k_+(d-x_w)] & u_0\sin[k_-(d-x_w)]
		\end{array}\right)
\end{aligned}
\end{equation}
\end{widetext}
In Eq.~\eqref{matrix}, we have approximated $k_\pm=k_F\varphi$ outside of the trigonometric functions while keeping $k_\pm=k_F\varphi\pm i\Omega/(v_F\varphi)$ inside. Taking the determinant of the matrix in Eq.~\eqref{matrix}, we find a solvability condition given by
\begin{equation} \label{soltemp1}
\begin{aligned}
0&=\Omega\varphi^2\sin(k_+d)\sin(k_-d)\biggl\{E^2-\xi_k^2-\frac{\gamma^2}{\varphi^2}\beta_+\beta_- \\
	&+\frac{\gamma}{\varphi}\left[\xi_k(\beta_++\beta_-)+\frac{E^2}{i\Omega}(\beta_+-\beta_-)\right]\biggr\},
\end{aligned}
\end{equation}
where we define $\beta_\pm=\{\cos[k_\pm(d-2x_w)]-\cos(k_\pm d)\}/\sin(k_\pm d)$. Dividing Eq.~\eqref{soltemp1} through by the common factor $\Omega\varphi^2\sin(k_+d)\sin(k_-d)$ and rearranging, we obtain
\begin{equation}
\begin{aligned}
0&=E^2\left(1+\frac{\gamma}{\Omega\varphi}\,\text{Im}(\beta_+)\right)^2-\frac{\Delta^2\gamma^2}{\Omega^2\varphi^2}[\text{Im}(\beta_+)]^2 \\
	&-\left(\xi_k-\frac{\gamma}{\varphi}\,\text{Re}(\beta_+)\right)^2
\end{aligned}
\end{equation}
Defining the quantities $\Gamma=\{1+\gamma\,\text{Im}(\beta_+)/(\Omega\varphi)\}^{-1}$ and $\delta\mu=\gamma\,\text{Re}(\beta_+)/\varphi$, we arrive at the solvability condition presented in Eq.~\eqref{solvability}. Substituting $k_\pm=k_F\varphi\pm i\Omega/(v_F\varphi)$ into the expressions for $\Gamma$ and $\delta\mu$, we arrive at the definitions presented in Eq.~\eqref{effparams}.

We also note that the solvability condition of Eq.~\eqref{solvability} can be expressed as $\det[G_w^R(E,k_y)]^{-1}=0$, where
\begin{equation}
(G_w^R)^{-1}=\left(\begin{array}{cc}
	E/\Gamma-\xi_k+\delta\mu & -\Delta(1/\Gamma-1) \\
	-\Delta(1/\Gamma-1) & E/\Gamma+\xi_k-\delta\mu
	\end{array}\right)
\end{equation}
Noting that the bare retarded Green's function of the nanowire is given by $(G_0^R)^{-1}=E-\xi_k\tau_z+i0^+$, the full Green's function can be written as $(G_w^R)^{-1}=(G_0^R)^{-1}-\Sigma^R$ with self-energy
\begin{equation}
\Sigma^R=\left(\begin{array}{cc}
	E(1-1/\Gamma)-\delta\mu & \Delta(1/\Gamma-1) \\
	\Delta(1/\Gamma-1) & E(1-1/\Gamma)+\delta\mu
	\end{array}\right),
\end{equation}
as given in Eq.~\eqref{selfenergy}.

\section{Integrating out superconductor} \label{appB}
In this section, we show how to alternatively derive Eqs.~\eqref{effparams} and \eqref{selfenergy} by integrating out the superconductor. For completeness, we first go through the steps of performing the integration. We start with the same model as considered in the main text, expressed in terms of the Euclidean action. The action of the nanowire is given by
\begin{equation}
S_{NW}=\frac{1}{2}\int\frac{d\omega}{2\pi}\int\frac{dk_y}{2\pi}\,\Psi_w^\dagger(G_w^0)^{-1}\Psi_w,
\end{equation}
where $(G_w^0)^{-1}=i\omega-\xi_k\tau_z$ is the inverse Matsubara Green's function of the nanowire in the absence of tunneling ($\omega$ is a Matsubara frequency) and $\Psi_w=[\psi_\uparrow(k_y,\omega),\psi_\downarrow^\dagger(-k_y,-\omega)]^T$ is a Heisenberg spinor field describing states in the nanowire. The action of the superconductor is given by
\begin{equation}
S_\text{BCS}=\frac{1}{2}\int\frac{d\omega}{2\pi}\int\frac{dk_y}{2\pi}\int_0^ddx\,\Psi_s^\dagger(i\omega-\mathcal{H}_\text{BCS})\Psi_s,
\end{equation}
where $\mathcal{H}_\text{BCS}$ is as defined below Eq.~\eqref{Hs} and $\Psi_s=[\eta_\uparrow(x,k_y,\omega),\eta^\dagger_\downarrow(x,-k_y,-\omega)]$ is a spinor field describing states in the superconductor. The tunneling action, which couples the nanowire to the superconductor at $x=x_w$, is taken to be
\begin{equation}
S_t=-\frac{t}{2}\int\frac{d\omega}{2\pi}\int\frac{dk_y}{2\pi}\int_0^ddx[\Psi_w^\dagger\tau_z\Psi_s\delta(x-x_w)+H.c.].
\end{equation}
The path integral representation of the partition function is then given by
\begin{equation}
\mathcal{Z}=\int D[\Psi_w^\dagger,\Psi_w]\int D[\Psi_s^\dagger,\Psi_s]e^{-S_w-S_\text{BCS}-S_t}.
\end{equation}
In the exponential, we rewrite
\begin{widetext}
\begin{equation} \label{shift}
\begin{aligned}
S_\text{BCS}+S_t&=\frac{1}{2}\int\frac{d\omega}{2\pi}\int\frac{dk_y}{2\pi}\biggl\{\int_0^ddx\biggl[\Psi_s^\dagger-t\Psi_w^\dagger \tau_zG_s^0(x_w,x)\biggr](i\omega-\mathcal{H}_\text{BCS})\biggl[\Psi_s-tG_s^0(x,x_w)\tau_z\Psi_w\biggr]-t^2\Psi_w^\dagger\tau_z G_s^0(x_w,x_w)\tau_z\Psi_w\biggr\}.
\end{aligned}
\end{equation}
\end{widetext}
In Eq.~\eqref{shift}, we introduce a function $G_s^0(x,x')$ that must satisfy $(i\omega-\mathcal{H}_\text{BCS})G_s^0(x,x')=G_s^0(x',x)(i\omega-\mathcal{H}_\text{BCS})=\delta(x-x')$; i.e. $G_s^0(x,x')$ corresponds to the Green's function of the superconductor in the absence of tunneling. Evaluating the Gaussian path integral over superconducting fermions, we obtain an effective action describing the nanowire given by
\begin{equation}
S_\text{eff}=\frac{1}{2}\int\frac{d\omega}{2\pi}\int\frac{dk_y}{2\pi}\Psi^\dagger_w\left[(G_w^0)^{-1}-\Sigma\right]\Psi_w,
\end{equation}
with the self-energy given by
\begin{equation} \label{selfenergyintout}
\Sigma=t^2\tau_zG_s^0(x_w,x_w)\tau_z.
\end{equation}

To explicitly evaluate the self-energy, we must choose the appropriate bare Green's function $G_s^0(x,x')$ for the geometry under consideration. For our purposes, we evaluate the self-energy using the Green's function of a finite-sized superconductor satisfying vanishing boundary conditions at $x=0$ and $x=d$. The bare Green's function must satisfy the equation
\begin{equation} \label{diffeq}
\left[i\omega+\left(\frac{\partial_x^2}{2m_s}-\frac{k_y^2}{2m_s}+\mu_s\right)\tau_z-\Delta\tau_x\right]G_s^0(x,x')=\delta(x-x').
\end{equation}
The solution to Eq.~\eqref{diffeq} can be written as the sum of a homogeneous solution $G_h(x,x')$ and a particular solution $G_p(x-x')$,
\begin{equation} \label{diffeqsol}
G_s^0(x,x')=G_h(x,x')+G_p(x-x'),
\end{equation}
with the particular solution corresponding to the bulk superconducting Green's function. We determine the bulk Green's function in real space by Fourier transformation. Defining $\xi_{ks}=(k_x^2+k_y^2)/2m_s-\mu_s$, we have
\begin{equation} \label{bulkG}
\begin{aligned}
G_p(x-x')&=-\int\frac{dk_x}{2\pi}\frac{i\omega+\xi_{ks}\tau_z+\Delta\tau_x}{\Delta^2+\xi_{ks}^2+\omega^2}e^{ik_x(x-x')} \\
	&=-\frac{1}{v_F\Omega\varphi}\biggl[(i\omega+\Delta\tau_x)\cos(\zeta|x-x'|) \\
	&-\Omega\tau_z\sin(\zeta|x-x'|)\biggr]e^{-\chi|x-x'|},
\end{aligned}
\end{equation}
where, as we have done throughout, we replace $k_\pm=\zeta=k_F\varphi$ outside of the exponentials while keeping $k_\pm=\zeta\pm i\chi=k_F\varphi\pm i\Omega/(v_F\varphi)$ inside of the exponentials (in Matsubara frequency space, $\Omega=\sqrt{\Delta^2+\omega^2}$). The homogeneous solution is given by
\begin{equation} \label{homogeneousG}
\begin{aligned}
G_h(x,x')&=(i\omega+\Delta\tau_x+i\Omega\tau_z)\left[c_1e^{ik_+x}+c_2e^{-ik_+x}\right] \\
	&+(i\omega+\Delta\tau_x-i\Omega\tau_z)\left[c_3e^{ik_-x}+c_4e^{-ik_-x}\right].
\end{aligned}
\end{equation}
The unknown coefficients, which are functions of the coordinate $x'$, are determined by imposing the boundary conditions $G_s^0(0,x')=G_s^0(d,x')=0$. Solving the boundary conditions gives
\begin{equation} \label{coeffs}
\begin{aligned}
c_1(x')&=\frac{\sin[k_+(d-x')]}{2v_F\Omega\varphi\sin(k_+d)}, \\
c_2(x')&=\frac{1}{2v_F\Omega\varphi}[i+\cot(k_+d)]\sin(k_+x'), \\
c_3(x')&=\frac{1}{2v_F\Omega\varphi}[-i+\cot(k_-d)]\sin(k_-x'), \\
c_4(x')&=\frac{\sin[k_-(d-x')]}{2v_F\Omega\varphi\sin(k_-d)}.
\end{aligned}
\end{equation}
Substituting Eqs.~\eqref{bulkG}-\eqref{coeffs} into Eq.~\eqref{diffeqsol} and setting $x=x'=x_w$, we find the bare Green's function
\vfill
\begin{widetext}
\begin{equation} \label{bareG}
\begin{aligned}
G_s^0(x_w,x_w)&=-\frac{1}{v_F\Omega\varphi[\cosh(2\chi d)-\cos(2\zeta d)]}\biggl\{(i\omega+\Delta\tau_x)\bigl\{\sinh(2\chi d)-\cos(2k\zeta x_w)\sinh[2\chi(d-x_w)] \\
	&-\cos[2\zeta(d-x_w)]\sinh(2\chi x_w)\bigr\}-\Omega\tau_z\bigl\{\sin(2\zeta d)-\sin(2\zeta x_w)\cosh[2\chi(d-x_w)]-\sin[2\zeta(d-x_w)]\cosh(2\chi x_w)\bigr\}\biggr\}.
\end{aligned}
\end{equation}
\end{widetext}
Substituting Eq.~\eqref{bareG} into Eq.~\eqref{selfenergyintout} and defining $\Gamma$ and $\delta\mu$ as in Eq.~\eqref{effparams}, we obtain a self-energy given by (recall $\gamma=t^2/v_F$)
\begin{equation}
\Sigma=(\Delta\tau_x-i\omega)(1/\Gamma-1)-\delta\mu\,\tau_z.
\end{equation}
After analytic continuation, we reproduce the retarded self-energy given in Eq.~\eqref{selfenergy}.

We note that choosing the bare Green's function to be translationally invariant and equal to the bulk superconducting Green's function, as is typically done to describe proximitized nanowires, corresponds to the geometry of a nanowire coupled to the middle of an infinitely large superconductor. In this case, the bare Green's function is simply $G_s^0(x_w,x_w)=G_p(0)$, yielding a self-energy
\begin{equation}
\Sigma_\text{bulk}=-\frac{\gamma(i\omega-\Delta\tau_x)}{\varphi\sqrt{\Delta^2+\omega^2}}.
\end{equation}

\bibliography{bibStrongCoupling}

\begin{thebibliography}{54}
\bibitem [{com()}]{comment}%
  \BibitemOpen
  \href@noop {} {}\bibinfo {note} {Note that the Andreev bound states for which
  the $g$-factor is measured experimentally do not exhibit a renormalization by
  the superconductor as suggested by Eq.~\eqref{topcriterion}
  \cite{Kouwenhoven}. This should be expected because the Andreev bound states
  originate from the ends of the nanowire where the superconducting shell is
  absent (and therefore $\Gamma=1$).}\BibitemShut {Stop}%
\end{thebibliography}

\end{document}